# Optical frequency divider with division uncertainty at the $10^{-21}$ level

Yuan Yao,[1] Yanyi Jiang,[1,2,*] Hongfu Yu,[1] Zhiyi Bi,[1,2] and Longsheng Ma[1,2,*]

[1]State Key Laboratory of Precision Spectroscopy, East China Normal University, Shanghai 200062, China

[2]Collaborative Innovation Center of Extreme Optics, Shanxi University, Taiyuan, Shanxi 030006, China

[*]Corresponding author: yyjiang@phy.ecnu.edu.cn, lsma@phy.ecnu.edu.cn

Optical clocks with unprecedented accuracy of $10^{-18}$ will lead to innovations in many research areas. All the applications of optical clocks rely on the ability of precisely converting the frequency from one optical clock to another, or particularly to the frequencies in the fiber telecom band for long-distance transmission. Here, we report a low-noise, high precision optical frequency divider. It can realize accurate optical frequency conversion as well as enable precise measurement of optical frequency ratios. By comparing against the frequency ratio between the fundamental and the second harmonic of a 1064 nm laser rather than a second similar system, the optical frequency divider is demonstrated to have a frequency division instability of $6 \times 10^{-19}$ at 1 s and a fractional frequency division uncertainty of $1.4 \times 10^{-21}$, nearly three orders of magnitude better than the most accurate optical clocks. It allows optical clocks to be accessible to many precision measurement applications.

Recent progress on optical atomic clocks demonstrates that the fractional frequency instability and uncertainty of optical clocks have been reduced to the $10^{-18}$ level [1–4]. The unprecedented accuracy provided by optical clocks will lead to a revolution in science and technology [5]. Using optical clocks, scientists are able to search for possible variations of fundamental constants in laboratories by precisely measuring the frequency ratios of two different atomic transitions of optical clocks over time [6–8]. In relativistic geodesy, long-distance geopotential difference will be accurately measured by comparing the frequencies of remote optical clocks linked with optical fibers [5,9,10], where the frequencies of optical clocks have to be accurately converted to those in the fiber telecom band for long-distance transmission. In metrology, the *second* in the International System of Units (SI) will be redefined based on optical atomic clocks [11]. Frequency comparisons between optical clocks based on different atom species have to be performed in order to ascertain the agreement between optical clocks with uncertainty beyond the current SI second, as well as the frequency reproducibility of optical clocks [12]. Moreover, atomic and molecular spectroscopists hope that the frequency accuracy and high frequency stability of optical clocks can be transferred to a wide spectral range for high precision spectroscopy.

All those applications rely on accurate frequency ratio measurement between spectrally-separated optical clocks or frequency conversion of optical clocks. Optical frequency combs [13,14] are employed to link the frequencies of optical clocks. By synchronously counting the frequency of the repetition rate ($f_r$) and the carrier-envelope offset frequency ($f_0$) of the comb, as well as the beat frequencies between optical clocks and the nearby comb teeth relative to a common hydrogen maser, frequency ratios with statistical uncertainty at the $10^{-17}$ level are obtained [6,15]. Using synchronous counting with a hydrogen maser and the transfer oscillator scheme [16,17], the systematic instability in optical frequency ratio measurement is limited to $2.8 \times 10^{-16}$ at 1 s averaging time [18]. With the current experiments on ultrastable lasers which has a frequency instability of $10^{-17}$ and beyond in the near future [19–22], together with the efforts on using correlated atomic samples to overcome the standard quantum limit [23,24], the frequency stability of optical atomic clocks will set a new record, demanding an even more precise way to quickly determine the frequency ratios.

Here we report a low-noise optical frequency divider (OFD) in the visible and near infrared region, capable of linking all the present optical clocks. Using the OFD, the frequency ratios between optical clocks can be precisely measured without synchronous counting the beating frequencies between optical signals and a comb against a hydrogen maser. Moreover, OFD can also accurately convert the frequency from one optical clock with a preset division ratio to another, or to the frequencies in the fiber telecom band [25] or in the microwave region [26] for a wide range of applications. Particularly, using an OFD to convert the frequency of one high-performance clock laser to another at a different wavelength, it improves the frequency ratio measurement by partially cancelling out the laser frequency noise in synchronization operation of optical clocks [15]. By comparing against the frequency ratio between the fundamental and the second harmonic of a 1064 nm laser, the fractional instability of the divisor is demonstrated to be $6 \times 10^{-19}$ at 1 s and $2 \times 10^{-20}$ at 1000 s, two orders of magnitude better than the most stable lasers [19,27–30]. The fractional uncertainty of the divisor in optical frequency division is characterized to be $1.4 \times 10^{-21}$. It can support frequency division of the most stable lasers and the most accurate optical clocks in the world without degrading the performance, and enables precision measurements at the $10^{-21}$ level.

The output light frequency of an OFD directly relates to the input light frequency with a precise ratio $R$, $f_{out} = f_{in}/R$. The experimental schematic of the OFD is shown in Fig. 1. An output laser ($f_{out}$) is phase-locked to the input light ($f_{in}$) via an optical frequency comb. The frequency of the $N$th comb tooth is $f_N = Nf_r + f_0$, where $N$ is an integer. To reduce the comb frequency noise, the comb is optically-referenced to $f_{in}$ by phase-locking the $N_1$th comb tooth to $f_{in}$ and stabilizing $f_0$ to a stable radio frequency (Supplement 1). A beat signal between the input laser light ($f_{in}$) and a nearby comb tooth ($f_{N1}$) can be written as

$$f_{b1} = f_{in} - f_{N1} = f_{in} - f_0 - N_1 f_r. \tag{1}$$

Meanwhile, another beat signal between the output laser ($f_{out}$) and a nearby comb tooth ($f_{N2}$) can be written as

$$f_{b2} = f_{out} - f_{N2} = f_{out} - f_0 - N_2 f_r, \tag{2}$$

where $N_1$ and $N_2$ are integers associated with the particular comb teeth. By measuring $f_{in}$ and $f_{out}$ on a wave meter with an uncertainty of 100 MHz, together with the values of the repetition rate $f_r$ and $f_0$ of the comb measured on frequency counters, $N_1$ and $N_2$ can be accurately determined. For the convenience of precise tuning of the output laser frequency $f_{out}$ as well as $R$ of the OFD, $f_{b2}$ is mixed with a tunable RF signal $f_{tune}$ on a double-balanced mixer (DBM) to obtain the signal as

$$f_{b2}^* = f_{out} - f_0 - N_2 f_r - f_{tune}. \tag{3}$$

The residual frequency noise of $f_0$ is removed by mixing $f_{b1}$ and $f_{b2}^*$ with $f_0$ in DBMs. The signal of $f_0$ is detected by using a collinear self-referencing 1$f$-2$f$ set-up [31], in which the detected extra frequency noise of $f_0$ due to the light path fluctuation is negligible. The outputs of the DBMs ($f_{b1}^\#$ and $f_{b2}^\#$) are sent to two DDSs with divisors of $M_1$ and $M_2$, respectively. Usually the signal to noise ratio (SNR) of a signal input to a DDS is more than 30 dB in a resolution bandwidth (RBW) of 300 kHz. The divisors of the DDSs, $M_1$ and $M_2$, are chosen to satisfy $M_1/M_2 = N_1/N_2$ in order to make the error signal $\Delta$ free from $f_r$. The outputs of the DDSs are compared on a DBM to generate an error signal as

$$\Delta = \frac{f_{in} - N_1 f_r}{M_1} - \frac{f_{out} - f_{tune} - N_2 f_r}{M_2} = \frac{f_{in}}{M_1} - \frac{f_{out} - f_{tune}}{M_2}. \tag{4}$$

Then the error signal is sent to a servo to adjust the frequency of the output laser to make $\Delta = 0$. As a result, $f_{out} = (M_2/M_1)f_{in} + f_{tune}$. In many applications of optical atomic clocks, it requires not only coherence transfer but also precisely setting the ratio between optical frequencies.

In order to set the division ratio precisely, here $f_{tune}$ has to be related to $f_{in}$ only. To achieve this goal, the beat signals of $f_{b1}$ and $f_{b1}^*$ between $f_{in}$ and the two nearest teeth of the same comb are detected on a photo detector. $f_{b1}^* = f_{N1+1} - f_{in} = f_0 + (N_1 + 1)f_r - f_{in}$. After removing $f_0$ and $f_r$ using DBMs and DDSs, the resulting signal is directly derived from $f_{in}$ as $(1/K_1 - 1/K_2) \times f_{in}$, independent on $f_0$ and $f_r$. $K_1$ and $K_2$ are the divisors of the DDSs, which are chosen to satisfy $K_1/K_2 = N_1/(N_1 + 1)$. In addition, a DDS with the divisor of $K_3$ is used to synthesize a self-referenced time base signal $f_{time}$ at about 10 MHz, $f_{time} = f_{in}/k$. Using this time base, a RF tuning frequency $f_{tune}$ is generated from a RF synthesizer (RF SYN) as $f_{tune} = f_{in}/K$, here $K$ depends on $K_1$, $K_2$, $K_3$ and the frequency setting of the RF synthesizer. Benefitting from the self-referenced RF signal, $f_{out}$ is directly divided from $f_{in}$ with a ratio of $R = 1/(M_2/M_1 + 1/K)$. If $f_{tune}$ is set with a resolution of 1 μHz, the ratio $R$ can be set at the 21th decimal place to an arbitrary pre-determined value when both $f_{in}$ and $f_{out}$ are within the spectrum of the comb. Meanwhile, $R$ can be precisely tuned [32] by sweeping $f_{tune}$.

To characterize the performance of the OFD, we measured the divisor $R$ of the OFD against the frequency ratio between the fundamental and the second harmonic of a 1064 nm laser instead of comparing against a second similar OFD. The second harmonic generation can realize optical frequency conversion, however, it is based on a completely different working principle from that of the OFD.

The experimental diagram of measurement is shown in Fig. 2(a). We used the OFD to divide the frequency of a cavity-stabilized laser at 1064 nm ($f_{1064-1}$, Supplement 1) by $R_x$ to a 532 nm laser ($f_{532}$). Here, the divisor is set randomly as 0.500 000 053 261 644 522 938. The beat notes of $f_{b2}$ (between $f_{532}$ and the comb) and $f_{b1}$ (between $f_{1064-1}$ and the comb) for optical frequency dividing are separately detected on photo detectors. Also, $f_{532}$ is second-harmonic generated from another independent 1064 nm laser ($f_{1064-2}$) in a nonlinear crystal. We assume the frequency ratio between the fundamental ($f_{1064-2}$) and its second harmonic ($f_{532}$) is exactly 2, that is $R_{SH} = f_{532}/f_{1064-2} = 2$. Both the laser light of $f_{532}$ and $f_{1064-2}$ propagate collinearly and combine with the optically-referenced frequency comb light on a beam splitter (Supplement 1). Besides, all the optics are well sealed in a box to reduce the fluctuation of light path due to airflow turbulence.

A beat note ($f_b$) between $f_{1064-1}$ and $f_{1064-2}$ is detected with a SNR of 40 dB (RBW = 300 kHz). The frequency of $f_b$ is nearly 30 MHz determined by $R_x$. Then $f_b$ is mixed down with a RF signal of 30.1 MHz from a RF SYN to nearly 100 kHz, which is then filtered in a low pass filter and is counted on a frequency counter ($\Lambda$-type, gate time = 1 s). The time base at about 10 MHz ($10^7$ Hz) for the RF frequency counter and synthesizers is the self-referenced time base of $f_{time} = f_{1064-1}/k$, as shown in Fig. 1. Therefore, $f_b$ is related to $f_{1064-1}$ as

$$f_b = \frac{30.1 \times 10^6 - A}{10^7} \times \frac{f_{1064-1}}{k}, \quad (5)$$

here $A$ is the reading number on the counter. With

$$f_b = f_{1064-1} - f_{1064-2} = f_{1064-1} - \frac{f_{532}}{R_{SH}} = f_{1064-1} - \frac{f_{1064-1}}{R_{SH} R_x}, \quad (6)$$

the frequency ratio $R_x$ can be obtained as

$$R_{\text{x}} = \frac{1}{R_{\text{SH}}\left(1 - \dfrac{30.1 \times 10^6 - A}{k \times 10^7}\right)}.$$

(7)

The fractional instability of the measured $R_{\text{x}}$ is shown in Fig. 2(b) with red dots. The low instability of $6 \times 10^{-19}$ at 1 s averaging time benefits from both the optically-referenced frequency comb as well as elimination of light path fluctuation (Supplement 1). The measured instability is close to the instability of $4 \times 10^{-19}/\sqrt{\tau}$ ($\tau$ is the averaging time) when two 1064 nm lasers are phase-locked to each other without using an optical frequency comb (blue squares). It implies that during optical frequency division, the relative frequency ratio instability between the output and input of the OFD is more than two orders of magnitude better than that of the most stable lasers [19,27–30], adding a neglectable noise onto the output of the OFD. When the light path in this measurement is not optimized (the light beams of $f_{532}$ and $f_{1064\text{-}2}$ propagate separately before beating against the comb light), it is hard for the instability to average down below $10^{-19}$ mainly due to the effect of vibration and temperature fluctuation on tens of centimeter of free-space propagation, as shown with purple triangles in Fig. 2(b). We measured the value of $R_{\text{x}}$ on nine different days over fifteen days. In Supplement 1, Fig. S4 shows all the measured data of $R_{\text{x}}$ (total measurement time of 105,000 s) as well as its fractional instability, which can be reduced down to the $10^{-21}$ level.

Figure 2(c) shows the fractional uncertainty of $R_{\text{x}}$ deviated from the setting value. Each data is averaged over 5000 s measurement time. Using standard statistical methods, we combine 21 set of data to calculate the weighted mean to be $0.4 \times 10^{-21}$ and a weighted fractional uncertainty to be $1.4 \times 10^{-21}$, corresponding to a 99% confidence level determined from a $\chi^2$ analysis [33]. The division uncertainty induced by the OFD is three orders of magnitude better than the most accurate optical clocks [3,4].

The merit of the transfer oscillator scheme is the immunity to comb frequency noise. Therefore, it is not necessary to phase-lock the comb to $f_{1064\text{-}1}$. However, due to the limited response bandwidth of the DDSs, the fast varying signals sent to the DDSs affect the performance of the system. We measured the frequency instability of the beat signal $f_{\text{b}}$ between two 1064 nm lasers ($f_{1064\text{-}1}$ and $f_{1064\text{-}2}$), whose frequencies are linked by the OFD as $f_{1064\text{-}2} = f_{1064\text{-}1}/R$, when a comb tooth is phase-locked to $f_{1064\text{-}1}$ but with $f_0$ free running. The drift rate of $f_0$ is within 20 kHz/s. As shown in Fig. 2(b) with green diamonds, the fractional instability of $R$ degraded to $1 \times 10^{-18}$ at 1 s and $4 \times 10^{-18}$ at 10 s. Therefore, in the experimental setup an optically-referenced frequency comb is employed, in which $f_0$ and $f_r$ fluctuate within ± 1 mHz measured with a gate time of 1 s. An additional benefit for an optically-referenced frequency comb is that it ensures that all the signals stay within the bandwidth of RF filters.

For the current experimental setup, the divisor range of the OFD in the optical domain is 0.5-1 if the input laser is at 1 μm, determined by the spectrum of the optical frequency comb in the range of 0.5 μm - 1 μm. While in the microwave region, the divisor can be set to $>10^7$. For example, the time base at ~10 MHz used for the RF synthesizers and counters is synthesized from the input frequency of the OFD, and the divisor $k$ for this signal is nearly $3 \times 10^7$. With non-linear techniques such as second harmonic generation or optical frequency summing, the divisor of the OFD could be extended to 0.5-10 if the input laser is at 1 μm, covering the optical wavelength from 0.5 μm to 10 μm.

Benefiting from the optically-referenced frequency comb, the self-referenced time base for frequency tuning and counting, and careful elimination of the light path fluctuation, an optical frequency divider with uncertainty at the $10^{-21}$ level has been demonstrated, meeting many applications of the state-of-the-art optical clocks as well as the next generation of optical clocks [34,35]. The frequency instability of $6 \times 10^{-19}$ at 1 s in optical frequency division exceeds the stability of any optical oscillator demonstrated to date [19,27–30] by more than two orders of magnitude. With the optical frequency divider, we can accurately measure the frequency ratios between optical signals, and accurately convert one laser frequency to other laser frequencies at the same time. We expect that this type of optical frequency divider will be instrumental in precision measurements.

**Funding**. National Natural Science Foundation of China (11334002 and 11374102); Shanghai Rising-Star Program (15QA1401900).

**Acknowledgment**. We thank C. Oates, Z. Lu, and J. Ye for valuable discussions and comments on this manuscript.

See Supplement 1 for supporting content.

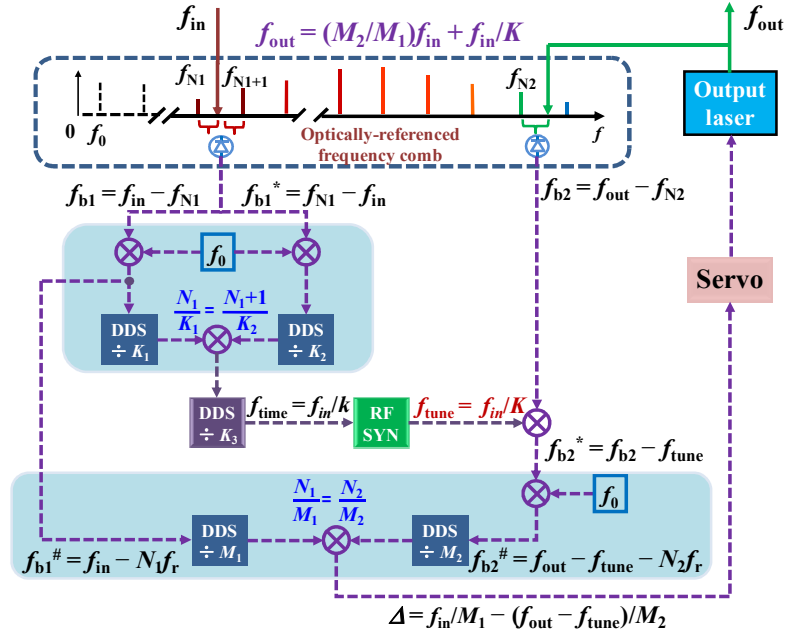

**Fig. 1.** Experimental realization of optical frequency divider (OFD). $f_{out}$ is related to $f_{in}$ with an optically-referenced frequency comb as $f_{out} = f_{in}/R$. The beating signals of $f_{b1}$ and $f_{b2}$ between $f_{in}$, $f_{out}$ and the nearby comb teeth are detected to generate an error signal $\Delta$ for phase-locking. The residual frequency noise of $f_0$ is subtracted from the beating signals on double balance mixers (DBM), and the residual frequency noise of $f_r$ is removed with direct digital synthesizers (DDS) and a DBM based on the transfer oscillator scheme. A tunable RF signal $f_{tune}$ synthesized from $f_{in}$ is used for precise adjustment of $f_{out}$ and $R$. RF SYN denotes RF synthesizer.

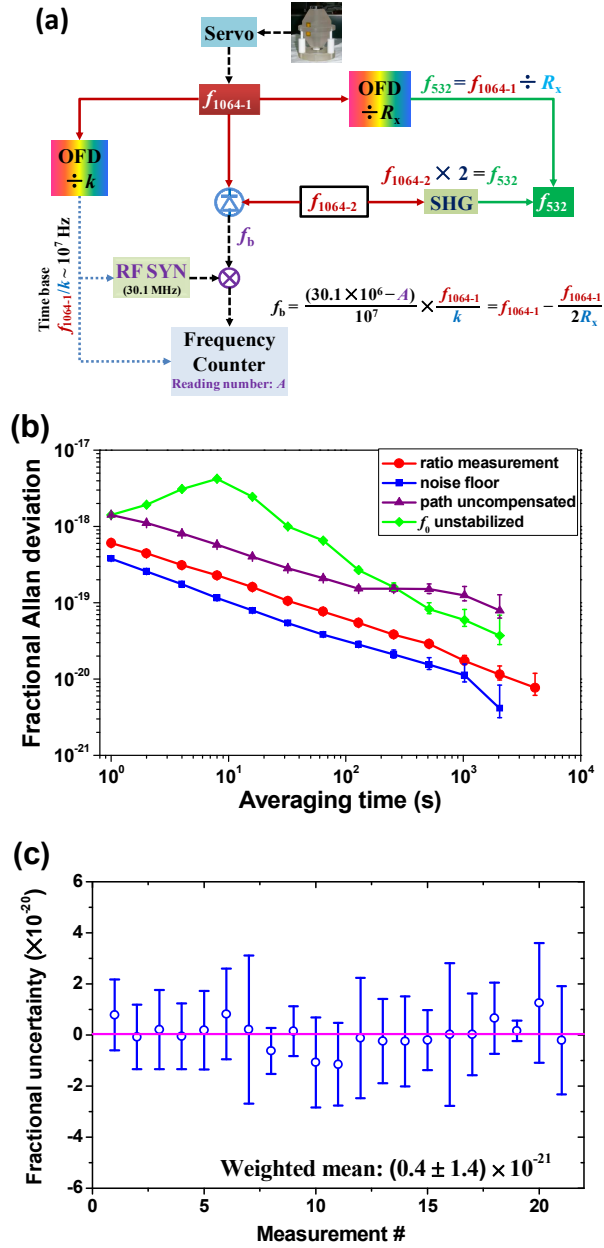

**Fig. 2.** Division noise and uncertainty measurement. (a) The frequency of a cavity-stabilized laser at 1064 nm ($f_{1064\text{-}1}$) is divided by $R_x$ to a 532 nm laser frequency ($f_{532}$) with an OFD. $f_{532}$ is second harmonic generated from an independent 1064 nm laser ($f_{1064\text{-}2}$). The uncertainty of $R_x$ is obtained by measuring the beat frequency ($f_b$) between $f_{1064\text{-}1}$ and $f_{1064\text{-}2}$ on a RF frequency counter with a self-referenced time base of $f_{1064\text{-}1}/k$. (b) Fractional Allan deviation of the frequency ratio between $f_{1064\text{-}1}$ and $f_{1064\text{-}2}$ when phase-locked to each other without the OFD (noise floor, blue squares), with the second harmonic generation (SHG) and the OFD (red dots). The purple triangles and green diamonds are the instability of the frequency ratio when light path

is not optimized and $f_0$ is free running accordingly. (c) Measured $R_x$ deviated from the setting value on nine different days over fifteen days. Each data point results from averaging over a roughly 5000 s measurement time. The error bar for each data point is the standard deviation of five mean values. Each mean value is averaged over 1000 s measurement time. The solid pink lines denote the mean value of the total 21 sets of data.

# Supplementary Materials:

## 1. OPTICALLY-REFERENCED FREQUENCY COMB

Figure S1 shows the experimental setup of the optically-referenced frequency comb. The optical frequency comb is based on a Ti:sapphire mode-locked femtosecond (fs) laser pumped by a 532 nm laser. It employs a six-mirror ring cavity with a repetition rate $f_r$ of 800 MHz. The output of the fs laser (average power of 600 mW, center wavelength at 810 nm) is coupled into a piece of photonic crystal fiber (PCF) for spectrum broadening. The PCF is sealed in an aluminum tube with end facet beam expansion, enabling long-time and robust running. A collinear self-referencing technique [1] is employed to detect the comb carrier-envelope offset frequency $f_0$ with a signal-to-noise ratio (SNR) of more than 50 dB in a resolution bandwidth (RBW) of 300 kHz. Using the collinear self-referencing technique, the comb light beams for generating the beating signal of $f_0$ propagate in a common path, resulting in a reduced detection frequency noise of $f_0$ due to the light path fluctuation. The signal of $f_0$ is phase-locked to a stable radio frequency (RF) signal by controlling the pumping power of the fs laser. After stabilized, $f_0$ fluctuates within $\pm 1$ mHz measured with a gate time of 1 s. Light from a cavity-stabilized 1064 nm laser ($f_{1064-1}$) co-propagates with the comb light through the PCF to a photo detector [2], eliminating the light-path-fluctuation-induced frequency noise. The beat note of $f_{b1}$ between $f_{1064-1}$ and the comb is detected with a SNR of more than 40 dB (RBW: 300 kHz). Then $f_{b1}$ is phase-locked to a stable RF synthesizer by controlling the cavity length of the fs laser via two piezo transducers (PZT) attached to cavity mirrors (fast and slow servo). As a result, each comb tooth has a frequency instability of $1.2 \times 10^{-15}$ at 1–40 s averaging time, limited by $f_{1064-1}$.

## 2. CAVITY-STAILIZED LASER AT 1064 NM

An Nd:YAG laser at 1064 nm is frequency-stabilized to an ultra-stable optical reference cavity using the Pound-Drever-Hall (PDH) technique. The reference cavity is 7.75 cm long and is vertically mounted to suppress the sensitivity to vibration. The cavity finesse is measured to be $\sim 2 \times 10^5$, corresponding to a cavity linewidth of ~10 kHz. The zero-expansion temperature of the reference cavity is measured to be ~22.87 °C. The cavity is held in a gold-coated copper thermal shield, which is enclosed in a vacuum chamber temperature-stabilized at 22.87 °C with sub-mK temperature instability. The vacuum chamber as well as some optical components for the PDH technique is mounted on a passive vibration isolation platform, which resides in a home-made acoustic isolation chamber with an acoustic attenuation of more than 20 dB. The acoustic isolation chamber is temperature-stabilized with a fluctuation of <70 mK. Laser light is transferred from the laser source to the reference cavity and to the comb through single-mode optical fibers with fiber noise compensation systems. Frequency stabilization is accomplished by controlling a PZT bonded on the monolithic Nd:YAG crystal of the laser to keep the laser frequency on resonance with the reference cavity. By comparing against a second similar laser system, the laser linewidth and frequency instability are measured to be 0.6 Hz and $1.2 \times 10^{-15}$ at 1–40 s averaging time, respectively.

## 3. TESTING THE IMMUNITY OF COMB FRQUENCY NOISE

If one of the divisors of the DDSs ($K_1$, $K_2$, $M_1$, $M_2$) is set incorrectly, we can easily observe the error by counting the beat frequency between two lasers linked with the optical frequency divider when tuning $f_0$ or $f_r$.

Figure S3 shows the beat frequency between two 1064 nm lasers linked by OFDs fluctuates over time when $f_0$ sweeps 3 kHz with a period of 10 s on purpose. $f_{1064\text{-}1}$ is firstly divided by $R_1$ to the frequency of a Ti:sapphire cw laser with an OFD, and then the frequency of the Ti:sapphire cw laser is divided by $R_2$ to $f_{1064\text{-}2}$ with another OFD based on the same comb. If the divisors of the DDSs for subtraction $f_r$ in one of the OFDs are set incorrectly, for example, $M_1/M_2 = (N_1+1)/N_2$, the beat frequency between the 1064 nm lasers fluctuates with a period of 10 s as well, as shown with the blue line in Fig. S3(b). Since one comb tooth of the optically-referenced frequency comb is frequency-stabilized to $f_{1064\text{-}1}$, $f_r$ changes when sweeping $f_0$. The imperfect subtraction of $f_r$ leads to the frequency fluctuation of $f_{1064\text{-}2}$. Moreover, it causes the laser to lose phase lock frequently. When the divisors of the DDSs are set correctly, the beat frequency between two lasers are more stable, as shown in Fig. S3(a).

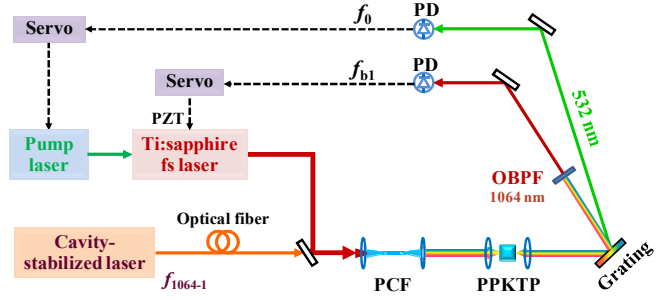

**Fig. S1.** Experimental setup for optically-referenced frequency comb. fs laser, femtosecond laser. PCF, photonic crystal fiber. PPKTP, periodically poled KTP. PD, photo detector. OBPF, optical band pass filter. $f_{b1}$, the beat note between $f_{1064\text{-}1}$ and the comb. $f_0$, the carrier-envelope offset frequency of the comb.

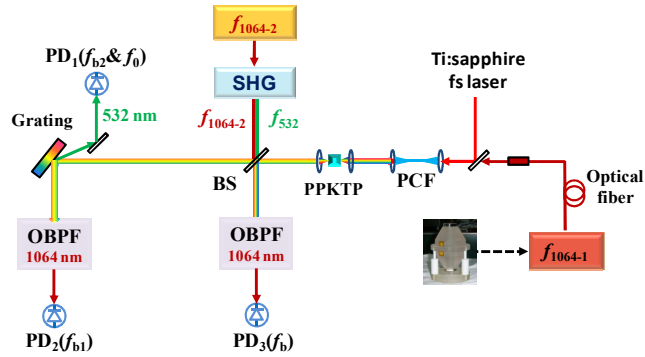

**Fig. S2.** Experimental setup for division noise and uncertainty measurement. fs laser, femtosecond laser. PCF, photonic crystal fiber. PPKTP, periodically poled KTP. SHG, second harmonic generation. BS, beam splitter. PD, photodiode. OBPF, optical band pass filter. $f_{b1}$, the beat note between $f_{1064\text{-}1}$ and the comb. $f_{b2}$, the beat note between $f_{532}$ and the comb. $f_0$, the carrier-envelope offset frequency of the comb. $f_b$, the beat note between $f_{1064\text{-}1}$ and $f_{1064\text{-}2}$.

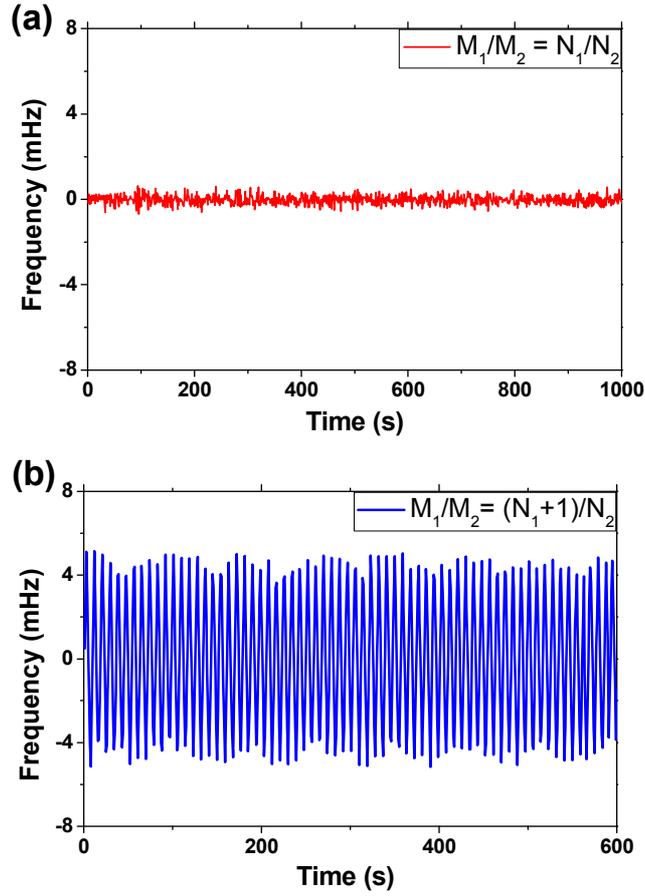

**Fig. S3.** Testing the immunity of comb frequency noise. $f_{1064\text{-}1}$ is firstly divided by $R_1$ to the frequency of a Ti:sapphire cw laser with an OFD, and then the frequency of the Ti:sapphire cw laser is divided by $R_2$ to $f_{1064\text{-}2}$ with another OFD based on the same comb. The figures show the fluctuation of the beat frequency between $f_{1064\text{-}1}$ and $f_{1064\text{-}2}$, when the comb offset frequency ($f_0$) fluctuates 3 kHz with a period of 10 s on purpose and the divisors of the DDSs for subtraction $f_r$ in one of division steps are set (a) correctly as $M_1/M_2 = N_1/N_2$ and (b) incorrectly as $M_1/M_2 = (N_1+1)/N_2$.

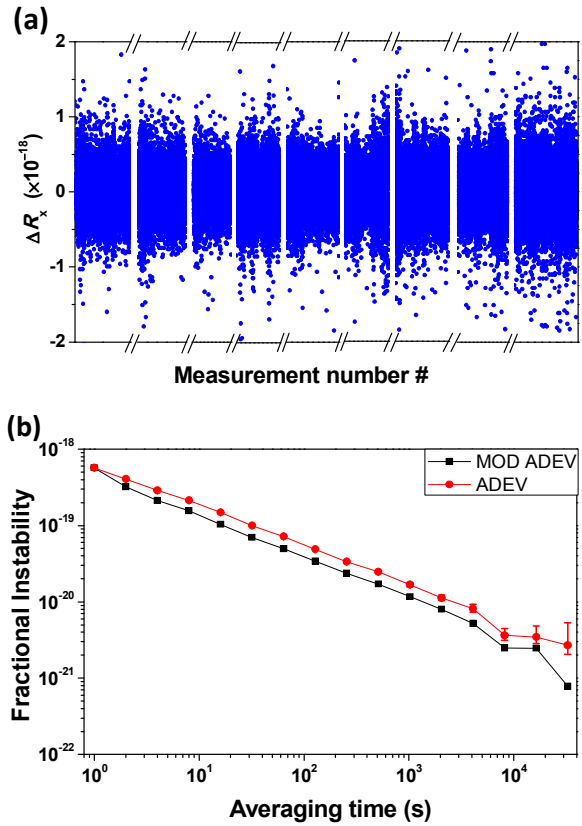

**Fig. S4.** Data for division noise and uncertainty measurement of an OFD when comparing against the frequency ratio between the fundamental and the second harmonic of $f_{1064\text{-}2}$. (a) All nearly 105000 data ($\Delta R_x$) of the measured ratio deviated from the set value. The value of $R_x$ is measured on nine different days. Each data is measured with a gate time of 1 s. (b) The Allan deviation (red dots) and the modified Allan deviation (black squares) of all the data.